\documentclass[12 pts]{article}
\usepackage{mathtools}
\usepackage{graphicx}
\usepackage{dcolumn}
\usepackage{bm}
\usepackage{amsmath}
\usepackage[T1]{fontenc}
\usepackage{authblk}
 \usepackage{booktabs}
 \usepackage{tikz}
 \usepackage{pgfplots}
 \usepackage{apacite}
\usepackage{caption}
\usepackage{subcaption}
\numberwithin{equation}{section} 
\usepackage{float}
\title{Defying Gravity: The Economic Effects of Social Distancing}
\author[1]{Alfredo D. Garcia}
\author[2]{Christopher Hartwell}
\author[3]{Martín Andrés Szybisz}
\affil[1]{Buenos Aires University, Faculty of Economic Science, e-mail garcia.alfredodaniel@gmail.com }
\affil[2]{International Management Institute, Zurich University of Applied Sciences; Department of International Management, Kozminski University, e-mail chartwell@kozminski.edu.pl }
\affil[3]{Buenos Aires University, Faculty of Economic Science, e-mail mszybisz@hotmail.com }
\begin{document}

\maketitle

\begin{abstract}
The COVID-19 pandemic has forced changes in production and especially in human interaction, with "social distancing" a standard prescription for slowing transmission of the disease. This paper examines the economic effects of social distancing at the aggregate level, weighing both the benefits and costs to prolonged distancing. Specifically we fashion a model of economic recovery when the productive capacity of factors of production is restricted by social distancing, building a system of equations where output growth and social distance changes are interdependent. The model attempts to show the complex interactions between output levels and social distancing, developing cycle paths for both variables. Ultimately, however, defying gravity via prolonged social distancing shows that a lower growth path is inevitable as a result.

\vspace{5mm}

\vspace{5mm}

\end{abstract}

\section{Introduction}\label{introduction}

It is a well-known maxim in physics tracing back to Newton \citeyear{newton1987philosophiae} that the force between two objects is proportional to the masses of these bodies but inversely proportional to the square of their geographical distance, suggesting that the proximity of an object is of paramount importance to its attraction to others. Indeed, the gravity model of trade \cite{isard1954location} is based precisely on this insight, predicting trade flows between countries to be higher (after controlling for trade costs) for countries which are more proximate. Extensions to the gravity model have also been employed for movements of peoples across borders and regions \cite{stouffer1940intervening, niedercorn1969economic, vanderkamp1977gravity}, using the size of various regions as attractants (an agglomeration effect) and the distance between specific regions as a deterrent to migration. While there have been some paradoxical findings from both of these literatures (see \cite{brun2005has} or Malmberg in \cite[p.~21-48]{hammar2021international}), the underlying premise of the benefits of proximity has been generally accepted. 

However, what occurs when distance is actually imposed, not a naturally occurring phenomenon of fixed geography but where economic actors are deliberately separated or forced further apart? Does the lesson of the gravity model hold, with artificial distance leading to less economic interaction and, subsequently, less output? The new reality of the COVID-19 pandemic has introduced precisely this possibility into economic life, as individual actors – in order to lessen the likelihood of catching the disease – are remaining “socially distanced” from each other; with an eye on keeping individuals at least 2 meters apart from each other, this distancing has been accomplished via a number of public policies and directives, including stay-at-home orders (both limited and more draconian), bans on large gatherings, and business closures \cite{abouk2021immediate}. 

In an economic sense, the effects of social distancing should be seen as a positive in the long-term, as the spread of a potentially fatal disease is halted \cite{sheridan2020social}. By halting the destruction of the labor stock and reducing the burdens on healthcare systems, a country can return to a long-term growth path \cite{bodenstein2021economic}. Seen in this light, social distancing acts as a collective good, imposing (hopefully) short-term restrictions and upfront costs in order to avoid larger costs in the long run. This would be amplified in certain specific conditions, where pre-existing attributes can make social distancing especially effective: for example, in countries where population has low density, where there are strong borders, or where there is either trust \cite{durante2021asocial} or social discipline which can make individuals adhere to social distancing more stringently. In these situations, social distancing can indeed be a one-time shock, and modelled accordingly \cite{mandel2020economic}. 

Unfortunately, the short-term consequences of social distancing may be substantial \cite{ashraf2020economic} and could even outweigh the longer-run costs of a protracted pandemic in situations where social distancing may not be a discrete event, leading to “waves” of contagion and, ultimately, output declines. In such an environment, there can be significant effects on poverty and inequality \cite{palomino2020wage} or the labor market \cite{gupta2020effects}. The distribution of effects is also not uniform and may fall heavily on specific sectors, in particular, industries which rely on turnover (such as hospitality and travel, see Gunay and Kurtulmus \citeyear{gunay2021covid} and Koren and Peto \citeyear{koren2020business}), which may find restrictions to be highly cumbersome \cite{de2020effect} or even impact the most vulnerable members of society (Mongey et al. 2020). Additionally, the uncertainty surrounding the ability of business to transact as normal (combined with other potentially economically destructive interventions by government) may deter long-term productive investment and increase risk aversion \cite{piccillo2020surprising}. Under these conditions, the collective good of social distancing becomes just another risk management decisions, sacrificed when other collective actions such as economic activity become necessary.
 
The purpose of this paper is to explore the economic ramifications of individual social distancing within a broader macroeconomy, capturing the complexity of the trade-offs between social distancing and economic activity. This is the first paper, to our knowledge, to incorporate the dynamics introduced during the COVID-19 pandemic in an attempt to understand the effect that social distancing might have on an economy. Building a model of production which accounts for the complex interaction between output level and social distancing, our results show that social distancing, if maintained, can result in lower output growth.

\section{The Literature on Distancing}\label{litreview}

Perhaps not surprisingly, given the extraordinary public health measures associated with “social distancing,” there is a sparse literature on the effects of physical distance amongst individuals on economic activity, as opposed to the vast literature on distance between countries. Indeed, most of the work done on social distancing over the past year has come from the public health and epidemiological literature, exploring the efficacy of distancing on stopping the spread of COVID-19. Papers such as Delen et al. \citeyear{delen2020no}, McGrail et al. \citeyear{mcgrail2020enacting}, and Park et al. \citeyear{park2020potential} provide data or modeling on the effectiveness of social distancing, and there appears to be near unanimity within the epidemiological and (especially) the public health literature on the desirability of social distancing in a pandemic.

Of the few economic analyses which have been produced, they too have focused on the economic effects of the disease and how distancing could, by hastening an end to the pandemic, have salutary economic benefits. Key in this vein is Greenstone and Nigham \citeyear{greenstone2020does}, who estimated that the lives saved from distancing, monetized using the US government’s value of a statistical life, could have a value of approximately US\$60,000 per household (as a working paper, the authors admit that there is no estimation of the corresponding costs of social distancing). A similar working paper from the NBER focuses instead on the cost side, coming to the conclusion that no social distancing would lead to COVID-related costs of US\$12,700 per person, while an optimal social distancing policy would reduce these costs to US\$8,100 per person; however, the “optimal policy” explored here would “curtail activity for decades or until a cure is found” \cite[p.~36]{farboodi2020internal}. Even though these costs are, like Greenstone and Nigham \citeyear{greenstone2020does}, linked to the costs associated with fatalities, the longer-run implications of curtailed activity over decade – and their likely astounding economic costs – are not explored. Thunström et al.\citeyear{thunstrom2020benefits}, on the other hand, attempt to combine both costs and benefits into a net tally and (using a slightly lower value of a statistical life, US\$10 million, than Greenstone and Nigham \citeyear{greenstone2020does}) find that the benefits of lived saved exceed the projected economic disruption of social distancing by US\$5.16 trillion.
  
Despite these papers focusing on the aggregate effects of social distancing, it is only a very small subset of these well-known papers which have examined the channels via which the economic cost of social distancing accrues, and which then attempt to array them against estimated benefits. Koren and Peto \citeyear{koren2020business}, focusing on industries which rely on face-to-face contact, find that sectors which use customer visits or have a high proportion of customer-facing workers have fared the worst due to social distancing measures; in terms of cost, the retail sector, forced to virtually shut during the first wave of the pandemic in the United States, would have required a 234\% wage subsidy in order to avoid unemployment losses. Similarly, Mongey et al. \citeyear{mongey2020workers}, using individual level data to show the most at-risk sectors, conclude that the costs of social distancing falls disproportionately on lower educated workers, those unable to work from home and those requiring high physical proximity, resulting in losses greater than those seen during the global financial crisis. Work from Barnett-Howell and Mubarak \citeyear{barnett2020benefits} echo this finding on an international level, showing that social distancing is both less effective and more destructive in poorer countries, increasing hunger, deprivation, and morbidity from other causes. Barrot et al. \citeyear{barrot2020sectoral} also expand this analysis beyond the most directly affected sectors to show how downstream industries in France are also affected, arriving at an estimate of the cost of social distancing at approximately 5.6\% of GDP. And, while attempting to show a positive view of social distancing as preventing larger and more protracted costs to an economy due to lost working hours and lives, Bodenstein et al. \citeyear{corsetti2020social} note that consumption holds up well despite social distancing, but at a cost of a lower capital stock over time. This certainly has longer-term ramifications for society, an intertemporal shift away from investment which can place an economy on a lower growth path. 
 
Another strand of the social distancing literature has focused on the drivers of compliance with social distancing as a contingency on their effectiveness. While not directly related to the idea of costs from social distancing policies, this work does reveal possible follow-on effects which can impact an economy. For example, research from Durante et al. \citeyear{durante2021asocial} shows that social distancing was utilized most on a voluntary basis in areas where there were high levels of “civic capital” (trust in others). In such a situation, cultural closeness could have mitigated the effects of social distance, although their work does not extend to the economic effects of this voluntary distancing. In a highly interesting study, Stockmaier et al.\citeyear{stockmaier2021infectious} also explore the voluntary social distancing behavior of animals and insects, finding that social distancing is limited where there is social or work pressure, and the need to provide and/or fend off other predators makes social distancing a luxury.

A final note is necessary here to bring us full circle: as noted at the beginning, the public health literature has concentrated on the efficacy of social distancing in reducing the spread of disease, but there are additional studies from this literature which hint at the broader costs from social distancing. For example, Venkatesh and Edirappuli \citeyear{venkatesh2020social} discuss the sometimes-sizeable effects which can come to mental health from social distancing, especially for those already at risk. Block et al. \citeyear{block2020social} recognize this issue and argue for a “strategic social network-based reduction of contact”, but as of yet the nuances of such alternatives to strict social distancing are in their infancy (meaning that the deleterious effects of actual isolation persist). There is unfortunately little understanding and no attempts (to our knowledge) to quantify these possible effects on the workforce and, thus, on economic activity.

\section{Formal presentation}\label{formalpresentation}

We have seen in section \ref{litreview} that the underlying effects of social distancing, and especially the channels by which these effects come to be, have been analyzed in the literature using their own form of distancing, i.e. being rather isolated and focusing on specific aspects of distancing rather than examining the issue holistically. Indeed, this approach has yielded some insights, as it has examined the long term benefits of preserving the labor force \cite{bodenstein2021economic, greenstone2020does} given the possible cost of distancing \cite{farboodi2020internal, greenstone2020does}, with the immediate economic cost also calculated \cite{koren2020business}. However, as the relations between the various attributes of social distancing are complex and may have differential economic impacts, this section attempts to present a more comprehensive view to better understand this complexity in an economic setting. 

In order to model these relationships, we first present a production function; the agent's decision exercise on how much labor he will supply subject to constrains which are adapted to the particular environment (social distancing and the risk of being out of the market due to the spread of the pandemic). Along the line of section \ref{introduction} these assumptions are very Newtonian, in the sense that their structure is very broadly used and is related to the forces acting upon the agent externally. 

The production function used in the model is:
\begin{equation}\label{productionfunctionK}
Y_{t}= F N\bar{K}
\end{equation}

by integrating constant capital $\bar{K}$ and the technological constant $F$ into a single coefficient $A$ for simplicity we get
\begin{equation}\label{productionfunction}
Y_{t}= A N
\end{equation}
where $Y_{t}$ represents output\footnote{We use output and income as synonyms in this paper.} of period $t$, $A$ is scale coefficient; whereas $N$ represents unconstrained production factors.
Given that a minimum production should be maintained we may subdivide \ref{productionfunction}  

\begin{equation}\label{productionfunctionmin}
Y_{t}= Y_{min}+A N
\end{equation}

\subsection{Budget restriction}

With fixed $\bar{K}$ and $A$, we may approximate

\begin{equation}\label{approxYN}
\frac{\dot{Y}}{Y}\approx\frac{\dot{N}}{N}
\end{equation}

Further, in view of \ref{approxYN}, we assume that the utilization proportion of the production factor $N$ develops in the following way;

\begin{equation}\label{budgetN}
N_{t+1}= N_{t}+\left[ gN_{t}+ \mu (-d)N_{t}\right] 
\end{equation}

where $g$ represents the growth rate of utilization of $N$ ( and therefore production output) and $\mu$ is a parameter defined in subsection \ref{parametermu}; while $d$ represents social distancing.

With the following restriction to production with capital fixed in the analysis period

\begin{equation}\label{budget}
Y_{t+1}= Y_{t}+\left[ gY_{t}+ \mu (-d)Y_{t}\right] 
\end{equation}

leading in continuous time to 

\begin{equation}\label{budgetc}
\frac{\dot{Y}}{Y}=\left(  g+ \mu (-d)\right)  
\end{equation}

Equation \ref{budgetc} shows the effect of social distancing as imposing a restriction over dynamic paths of output $Y$. The factor $g$ is a growth rate which may assume modest positive values in normal times (usually scaled as no greater that 0.1 per year), whereas in exceptional times or a crisis may assume negative values.

\subsection{Policy and structural parameters}\label{parametermu}

The factor $\mu$ of equation \ref{budgetc} is set as a parameter. The greater social distancing $d$ the more impact over output $Y$ will occur, as indicated by $\mu$. The higher $Y$ or $d$, the higher the restraint of the growth of $Y$. The intuition is that the higher income $Y$ is, the more possibilities exist to restrain or cut production as shown in section \ref{introduction}.

While the general direction is set by the relations of equation \ref{budgetc}, the precise quantitative impact depend on the set of complex economic policies\footnote{Among other drives as seen in section \ref{litreview}.} put forward. This is the justification to leave $\mu$ as a parameter with some degree of freedom. It is constrained by policies setters and by the reaction of the system and determined by the technological relation of the production function. For example, if a policy of closing borders is set, the impact on output is given by the existent technology and sector breakdown; i.e. if the tourism branch constitute a huge proportion of the economy $\mu$ will be greater.

As a concrete example of the problems involved in setting the value of $\mu$, consider the (extreme) case of an economy whose production consist almost entirely of the output of the software sector. In this case, it is even conceivable to assume that the second term of the RHS of equation \ref{budgetc} may be positive. 

Another important point of the model is that it is able to deal with the acceleration in the technological change in some labor markets routines. When the working force mobility is replaced by home working and teleworking, the output recovering may be faster and higher.

\subsection{Response to distancing}

For the possibility to derive the response of the agents as an individual behavior\footnote{It is important to note that this approach is not the only possibility, as cultural factors may be taken into account also as base for the proportion in which the effectiveness of $d$ decays.}, we take

\begin{equation}\label{utility}
U=U(C_{t},1-N_{S})
\end{equation}
 where $U$ represent utility and $C$ consumption. In this context, we interpret $1-N_{S}$ not as leisure, but as tolerance to social distancing $d$. 
which is subject to
\begin{equation}\label{wagerestriction}
P_{t}C_{t}=W_{t}N_{t}
\end{equation}
with $W$ standing for Wage and $P$ for prices.

Upon introducing the definition of consumption form equation \ref{wagerestriction} to equation \ref{utility} we obtain

\begin{equation}\label{utilityconsumtion}
U=U\left( \frac{W_{t}N_{t}}{P_{t}},1-N_{S}\right) 
\end{equation}

and by differentiating completely we obtain the first order condition for utility maximization for the agent

\begin{equation}\label{utilityconsumtionfirst}
\dfrac{dU}{dN}=\frac{W}{P}-U_{1-N_{S}} =0
\end{equation}
where the first term of the RHS measures the marginal benefit of the use of the unit of labor whereas the second term refers to the cost, which we assume 

\begin{equation}\label{utilityconsumtionfirstassumption}
U_{1-N_{S}} =\tau d
\end{equation}

By taking $N_{S}=N_{D}$ which is equivalent to state that factor's $N$ demand is enough at the wage rate $\frac{W}{P}$ to employ $N_{S}$.

So, social distancing $d$ will decline as

\begin{equation}\label{utilitydsimple}
\dot{d} =-\tau d
\end{equation}

The intuition behind equation \ref{utilitydsimple} is that, with time, wage earners will be willing to work for a lower wage rate. This not necessarily implies a lower real wage rate for each individual, it is conceivable that the wage rate becomes lower due to various type of risk involved; such as fines, difficulties in transportation and social critic among others. Form equation \ref{utilitydsimple} we may see that with time, owners of factor $N$ will tend to accept lower retributions because of most urgent necessities. The last effect operates simultaneously (is the counterpart of) with a lesser tolerance of social distancing $d$. 
Therefore we expect that $\tau$ will take positive values, less than one taking account of a progressive decay of the effectiveness of $d$.

\subsection{Susceptibility}

Now let us formalize the causes of growing social distancing $d$. This measure is imposed to protect factor $N$ by limiting social interaction. By doing so it reduces also output $Y$ by assumption \ref{approxYN}. For simplicity we assume that the growth of $d$ can be expressed with a parameter $\rho$. 

To explain parameter $\rho$ we may have in mind the effect of the infection by referring to the simple Susceptible-Infectious-Recovered (SIR) model:

\begin{equation}\label{Sdot}
\dot{\Omega}=a\Omega\Gamma
\end{equation}

\begin{equation}\label{Idot}
\dot{\Gamma}=a\Omega\Gamma-b\Gamma
\end{equation}
\begin{equation}\label{Rdot}
\dot{\Lambda}=b\Gamma
\end{equation}

where $\Omega$ is the fraction of susceptible individuals, $\Gamma$ is the fraction of infectious individuals, and $\Lambda$ is the fraction of recovered individuals\footnote{Where we suppose that all these proportions $\Omega,\Gamma,\Lambda$ are related to the unrestricted production factor $N$ and by equation \ref{approxYN} to output $Y$.}; $a$ is the transmission rate per infectious individual, and $b$ is the recovery rate. 

Linearizing about the disease-free equilibrium (DFE)

\begin{equation}\label{Idotl}
\dot{\Gamma}=(a-b)\Gamma
\end{equation}

Thus, if $a-b>0$, then $\Gamma$ grows exponentially around the DFE. 

By \ref{Idotl} we may approximate $\rho$ by the following equation:

\begin{equation}\label{rho}
\rho\approx\Phi(a-b)
\end{equation}

So, $\rho$ will be defined as function $\Phi$ of the rate of infection $a$ less recovered cases $b$. In this way we have no restrictions over $\rho$ allowing changes in a wide range to see the impact on activity and output. 

The variable $d$ will change as a function of the impact that $\rho$ (net of recovered) has on a given level of output.

Hence, social distancing $d$ will grow in a proportion $\rho$ respect to factor $N$ (which is susceptible to lose his productions capabilities) and therefore by assumption \ref{approxYN} to output $Y$.

In short, social distancing $d$ will be growing in proportion $\rho$ respect to income $Y$. 

This would lead to adjust equation \ref{utilitydsimple} with an additional term 
\begin{equation}\label{utilitydsimplecomplete}
\frac{\dot{d}}{d} =-\tau+\rho Y
\end{equation}

\subsection{The system of equations}\label{systemequations}

If we put together equations \ref{budgetc} and \ref{utilitydsimplecomplete} we obtain the system

\begin{eqnarray}
\frac{\dot{Y}}{Y} &=&g -\mu d \nonumber\\
\frac{\dot{d}}{d}&=& -\tau + \rho Y, \nonumber\\
\label{system}
\end{eqnarray}

This system defines the dynamic interactions between output and social distancing based on an unconstrained growth rate $g$, a parameter $\mu$, the reaction (decay of $d$) $\tau$ and $\rho$ which refers to the susceptibility of $d$ to income $Y$ or factor $N$ given the relation \ref{approxYN}. In particular, the assumptions are: a budget restriction given by equation \ref{budget}, a behavior assumption of agents via equation \ref{utilitydsimple} and the susceptibility of the factor to not being able to produce via equation \ref{rho}. The remaining factor $\mu$ is set as a policy and structural parameter which may vary with policies, economic structure or system responses to different scenarios; it refers to the effect that $d$ has over the variation of $Y$.

When all parameters have positive values, the point $\left( \dfrac{\tau}{\rho},\dfrac{g}{\mu}\right) $ is also the center around which the orbits eventually evolves, in this sense it is the second point of equilibrium apart from $(0,0)$.

We take variables $Y, d,\frac{W}{P}, N$ with dimension of indexes, whereas parameters $g,\mu,\tau,\rho$ are dimensionless. 

Note that if $d=0$ we obtain a simple case of income growing at his technical possibilities.

Under this framework, social distancing dynamics are lagging that of output\footnote{From equation \ref{system} we that $\dfrac{d\left( \frac{\dot{d}}{d}\right) }{dY}=\rho$.}, an increase in output means an increment of contacts of $N$. 

The dynamics of the model implies that some quantity adaptation in output is done; distancing necessarily triggers lower output\footnote{From equation \ref{system} we that $\dfrac{d\left( \frac{\dot{Y}}{Y}\right) }{dd}=-\mu$.}. 

Note that while the structure of the assumptions is very broadly used, we are not able to present deterministic results (or a range given some stochastic assumption). We present drivers of the parameters $g,\mu,\tau,\rho$ but we are not aware of models that have sufficient theoretical and predicting power to provide these type of quantification. The alternative is to simulate some paths.

\section{Simulations and results}\label{SR}

To avoid the problems that simple binary presentations of benefit and costs may suggest\footnote{See the rationale at the beginning of section \ref{formalpresentation}.} (i.e. health vs output) in terms of actions that they may trigger; we work with a more complete set of parameters which we presented in section \ref{formalpresentation}. 

As the interrelation between the different elements are extremely complex, in our attempt to get a sense of these combined dynamics we use some basic simulations (see subsection \ref{systemequations}).

We run dynamic paths\footnote{Using Maxima for calculations.}; two figures are presented, one for the space of output and distancing and another for the temporal evolution of both variables. We present the parameters in annual scale and use $Y$ for Income, $d$ for distancing, $T$ for trajectories and $t$ for time in the figures. 

\subsection{Extreme cases}

First we present two extreme cases. In both, the short range of time does not allow to show the impact of contagion or distancing interactions with output making it difficult to match real economic conditions. This extremely short time convergence may explain why we are not able to see examples of this paths in practice.

\begin{figure}[H]
	\centering
	\begin{subfigure}[b]{0.4\textwidth}
		\begin{tikzpicture}[scale=0.6]
		\begin{axis}[
		xlabel=Output,ylabel=Distancing,
		title={ Dynamic Paths},
		domain=-0:1,
		ymax=1,
		ymin=0,
		xmax=1,
		xmin=0,
		view={0}{90},
		axis background/.style={fill=white},
		]
		\addplot3 [red, mark=none, thick=3, smooth] table[z=z,x=In,y=dis] {

			z	In	dis
			0   1       0.449204
			1	0.850059  0.502558	
			2	0.588175  0.588175
			3   0.240763  0.475939
			6   0.0996185  0.183245
			8   0.0829129  0.0829129
			11   0.0776115 0.0247946
			15  0.0864143  0.00498826

		};
		\addlegendentry{T}

		\addplot3[->,red, thick, smooth] 
		table[z=z,x=I,y=d] {
			
			z	I	d
			6   0.0996185  0.183245
			8   0.0829129  0.0829129
			
		};

		\addplot3 [
		blue,-stealth,samples=15,
		quiver={
			u={(0.04*x-(1.1)*x*y)/(sqrt((0.04*x-(1.1)*x*y)^2+(-0.5*y+1*x*y)^2))},
			v={(-0.5*y+1*x*y)/(sqrt((0.04*x-(1.1)*x*y)^2+(-0.5*y+1*x*y)^2))},
			scale arrows=0.05,
		},
		] {0};
		\end{axis}
		\end{tikzpicture}
		
	\end{subfigure}
	~
	\begin{subfigure}[b]{0.4\textwidth}
		\begin{tikzpicture}[scale=0.6]
		\begin{axis}
		[
		xlabel=Time,ylabel=Distancing and Output,
		title={ Dynamic Paths},
		domain=-0:1,
		ymax=1,
		ymin=0,
		xmax=15,
		xmin=0,
		axis background/.style={fill=white},
		]
		
		\addplot [green, mark=none, thick=3, smooth] table[x=z,y=In] {
			
			z	In	dis  
			0   1       0.449204
			1	0.850059  0.502558	
			2	0.588175  0.588175
			3   0.240763  0.475939
			6   0.0996185  0.183245
			8   0.0829129  0.0829129
			11   0.0776115 0.0247946
			15  0.0864143  0.00498826

		};
		\addlegendentry{Y}
		\addplot [red, mark=none, thick=3, smooth] table[x=z,y=dis] {
			
			z	In	dis
			0   1       0.449204
			1	0.850059  0.502558	
			2	0.588175  0.588175
			3   0.240763  0.475939
			6   0.0996185  0.183245
			8   0.0829129  0.0829129
			11   0.0776115 0.0247946
			15  0.0864143  0.00498826
			
		};
		\addlegendentry{d}
		\end{axis}
		\end{tikzpicture}
	\end{subfigure}
	\caption{Case Recession}
	\label{phasedr}
\end{figure}
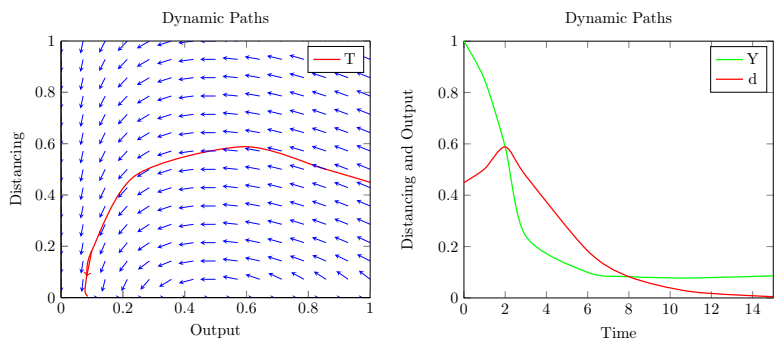

\begin{figure}[H]
	\centering
	\begin{subfigure}[b]{0.4\textwidth}
		\begin{tikzpicture}[scale=0.6]
		\begin{axis}[
		xlabel=Output,ylabel=Distancing,
		title={ Dynamic Paths},
		domain=-0:1,
		ymax=1,
		ymin=0,
		xmax=1,
		xmin=0,
		view={0}{90},
		axis background/.style={fill=white},
		]
		\addplot3 [red, mark=none, thick=3, smooth] table[z=z,x=In,y=dis] {
			
			z	In	dis
			0	0.15	1
			1	0.233862	0.63439
			4   0.5 0.25
			5   0.577171  0.18
			6   0.640992  0.139231
			8   0.795041  0.10402
			10  0.935886  0.0798122
			12 1.1        0.066608

		};
		\addlegendentry{T}

		\addplot3[->,red, thick, smooth] 
		table[z=z,x=I,y=d] {
			
			z	I	d
			0   0.577171  0.18
			0   0.640992  0.139231
			
		};

		\addplot3 [
		blue,-stealth,samples=15,
		quiver={
			u={(0.04*x-(-0.49)*x*y)/(sqrt((0.04*x-(-0.49)*x*y)^2+(-0.46*y+0.37*x*y)^2))},
			v={(-0.46*y+0.37*x*y)/(sqrt((0.04*x-(-0.49)*x*y)^2+(-0.46*y+0.37*x*y)^2))},
			scale arrows=0.05,
		},
		] {0};
		\end{axis}
		\end{tikzpicture}
		
	\end{subfigure}
	~
	\begin{subfigure}[b]{0.4\textwidth}
		\begin{tikzpicture}[scale=0.6]
		\begin{axis}
		[
		xlabel=Time,ylabel=Distancing and Output,
		title={ Dynamic Paths},
		domain=-0:1,
		ymax=1,
		ymin=0,
		xmax=10,
		xmin=0,
		axis background/.style={fill=white},
		]
		
		\addplot [green, mark=none, thick=3, smooth] table[x=z,y=In] {
			
			z	In	dis
			0	0.15	1
			1	0.233862	0.63439
			4   0.5 0.25
			5   0.577171  0.18
			6   0.640992  0.139231
			8   0.795041  0.10402
			10  0.935886  0.0798122
			12 1.1        0.066608

		};
		\addlegendentry{Y}
		\addplot [red, mark=none, thick=3, smooth] table[x=z,y=dis] {
			
			z	In	dis
			0	0.15	1
			1	0.233862	0.63439
			4   0.5 0.25
			5   0.577171  0.18
			6   0.640992  0.139231
			8   0.795041  0.10402
			10  0.935886  0.0798122
			12 1.1        0.066608
			
		};
		\addlegendentry{d}
		\end{axis}
		\end{tikzpicture}
	\end{subfigure}
	\caption{Case Recovery}
	\label{phaser}
\end{figure}
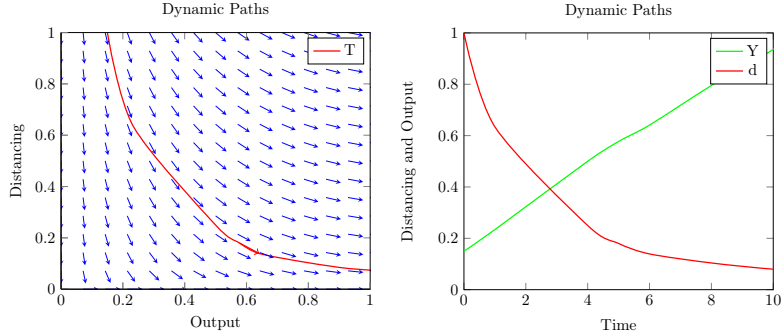

We may see a "$d$ dominant, recession" case in figure \ref{phasedr}\footnote{Passes at Y=0.25 and d=0.5, case $g=0.04,\mu=1.1,\tau=0.5,\rho=1$.}. For the recessive case the high value of reaction of the economy to distancing ($\mu=1.1$) does not allow the system to react to retain a cycle with income fluctuations.
We may label the trajectories of figure \ref{phaser} as case recovery\footnote{Passes at Y=0.5 and d=0.25, case $g=0.04,\mu=-0.49,\tau=0.46,\rho=0.37$.}. This would be a case when the economy is pushed by sector like software or logistics which experience growth given the need for his services.

\subsection{A path of changing parameters}

We show a sequence of cases where the first one represent the initial state of the system once distancing is imposed. In each case, the complete set of parameters is redefined, so the path is robust from a Lucas critique \cite{lucas1976econometric} perspective. 

In the case of figure \ref{phase1} we have high contagion with low distancing measures (agent have not learned how to protect themselves and distancing measures are at a low level)\footnote{Case $g=0.04,\mu=0.48,\tau=0.79,\rho=2$ that may pass at Y=0.3 and d=0.25.}.

\begin{figure}[H]
	\centering
	\begin{subfigure}[b]{0.4\textwidth}
\begin{tikzpicture}[scale=0.6]
\begin{axis}[
xlabel=Output,ylabel=Distancing,
title={ Dynamic Paths},
domain=-0:1,
ymax=1,
ymin=0,
xmax=1,
xmin=0,
view={0}{90},
axis background/.style={fill=white},
]
\addplot3 [red, mark=none, thick=3, smooth] table[z=z,x=In,y=dis] {
	
	z	In	dis
	0	0.3	0.25
	4	0.253668	0.110622
	7	0.264671	0.0446
	10	0.284478	0.02259390
	20  0.405516    0.00718897
	30  0.557355    0.0424002
	32  0.574971    0.09744178
	35  0.528756    0.214055
	38  0.390112    0.330692
	41	0.3	0.25
	45	0.253668	0.110622
	48	0.264671	0.0446
	51	0.284478	0.02259390

};
\addlegendentry{T}

\addplot3[->,red, thick, smooth] 
table[z=z,x=I,y=d] {
	
	z	I	d
	0	0.3	0.25
	0	0.253668	0.110622
	
};

\addplot3 [
blue,-stealth,samples=15,
quiver={
	u={(0.04*x-0.48*x*y)/(sqrt((0.04*x-0.48*x*y)^2+(-0.79*y+2*x*y)^2))},
	v={(-0.79*y+2*x*y)/(sqrt((0.04*x-0.48*x*y)^2+(-0.79*y+2*x*y)^2))},
	scale arrows=0.05,
},
] {0};
\end{axis}
\end{tikzpicture}
\end{subfigure}
 ~
\begin{subfigure}[b]{0.4\textwidth}
\begin{tikzpicture}[scale=0.6]
\begin{axis}
[
xlabel=Time,ylabel=Distancing and Output,
title={ Dynamic Paths},
domain=-0:1,
ymax=1,
ymin=0,
xmax=100,
xmin=0,
axis background/.style={fill=white},
]

\addplot [green, mark=none, thick=3, smooth] table[x=z,y=In] {

	z	In	dis
	0	0.3	0.25
	4	0.253668	0.110622
	7	0.264671	0.0446
	10	0.284478	0.02259390
	20  0.405516    0.00718897
	30  0.557355    0.0424002
	32  0.574971    0.09744178
	35  0.528756    0.214055
	38  0.390112    0.330692
	41	0.3	0.25
	45	0.253668	0.110622
	48	0.264671	0.0446
	51	0.284478	0.02259390
	61  0.405516    0.00718897
	71  0.557355    0.0424002
	73  0.574971    0.09744178
	76  0.528756    0.214055
	79  0.390112    0.330692
	82	0.3	0.25
	86	0.253668	0.110622
	89	0.264671	0.0446
	92	0.284478	0.02259390
	102  0.405516    0.00718897
	
};
\addlegendentry{Y}
\addplot [red, mark=none, thick=3, smooth] table[x=z,y=dis] {
	
	z	In	dis
	0	0.3	0.25
	4	0.253668	0.110622
	7	0.264671	0.0446
	10	0.284478	0.02259390
	20  0.405516    0.00718897
	30  0.557355    0.0424002
	32  0.574971    0.09744178
	35  0.528756    0.214055
	38  0.390112    0.330692
	41	0.3	0.25
	45	0.253668	0.110622
	48	0.264671	0.0446
	51	0.284478	0.02259390
	61  0.405516    0.00718897
	71  0.557355    0.0424002
	73  0.574971    0.09744178
	76  0.528756    0.214055
	79  0.390112    0.330692
	82	0.3	0.25
	86	0.253668	0.110622
	89	0.264671	0.0446
	92	0.284478	0.02259390
	102  0.405516    0.00718897
	
};
\addlegendentry{d}
\end{axis}
\end{tikzpicture}
\end{subfigure}
\caption{Phase 1}
\label{phase1}
\end{figure}

Once measures of distancing are imposed and as seen in section \ref{litreview} social disciple holds (lower $\tau$); output fluctuates at lower levels and contagion is somewhat smaller too\footnote{Case $g=0.04,\mu=0.29,\tau=0.46,\rho=1.75$ that may pass at Y=0.2 and d=0.2.}.

\begin{figure}[H]
	\centering
	\begin{subfigure}[b]{0.4\textwidth}
\begin{tikzpicture}[scale=0.6]
\begin{axis}[
xlabel=Output,ylabel=Distancing,
title={ Dynamic Paths},
domain=-0:1,
ymax=1,
ymin=0,
xmax=1,
xmin=0,
view={0}{90},
axis background/.style={fill=white},
]
\addplot3 [red, mark=none, thick=3, smooth] table[z=z,x=In,y=dis] {
	
	z	In	dis
	0	  0.2	0.2
	3     0.194249  0.145833	
    10    0.205252 0.066608	
	18    0.255869 0.0424002
	33    0.343897 0.154636
	40   0.295481  0.295481
	42   0.253668  0.313087
	49   0.2 0.2
	 52   0.194249  0.145833
	59   0.205252  0.066608 
	67    0.255869 0.0424002
	82    0.343897 0.154636	
	89   0.295481  0.295481
	91   0.253668  0.313087
	98   0.2 0.2
	101   0.194249  0.145833
};
\addlegendentry{T}

\addplot3[->,red, thick, smooth] 
table[z=z,x=I,y=d] {
	
	z	I	d
    0	0.2	0.2
    3   0.194249  0.145833
	
};

\addplot3 [
blue,-stealth,samples=15,
quiver={
	u={(0.04*x-0.29*x*y)/(sqrt((0.04*x-0.29*x*y)^2+(-0.46*y+1.75*x*y)^2))},
	v={(-0.46*y+1.75*x*y)/(sqrt((0.04*x-0.29*x*y)^2+(-0.46*y+1.75*x*y)^2))},
	scale arrows=0.05,
},
] {0};
\end{axis}
\end{tikzpicture}
\end{subfigure}
~
\begin{subfigure}[b]{0.4\textwidth}

\begin{tikzpicture}[scale=0.6]
\begin{axis}
[
xlabel=Time,ylabel=Distancing and Output,
title={ Dynamic Paths},
domain=-0:1,
ymax=1,
ymin=0,
xmax=100,
xmin=0,
axis background/.style={fill=white},
]

\addplot [green, mark=none, thick=3, smooth] table[x=z,y=In] {

	z	In	dis
	0	0.2	0.22
	3   0.194249  0.145833
    10    0.205252 0.066608
	18    0.255869 0.0424002
	33    0.343897 0.154636
	40   0.295481  0.295481
	42   0.253668  0.313087
	49   0.2 0.2
	 52   0.194249  0.145833
	59   0.205252  0.066608 
	67    0.255869 0.0424002
	82    0.343897 0.154636	
	89   0.295481  0.295481
	91   0.253668  0.313087
	98   0.2 0.2
	101   0.194249  0.145833
};
\addlegendentry{Y}
\addplot [red, mark=none, thick=3, smooth] table[x=z,y=dis] {
	
	z	In	dis
    0	0.2	0.2
    3   0.194249  0.145833
    10   0.205252  0.066608 
    18    0.255869 0.0424002
    33    0.343897 0.154636	
    40   0.295481  0.295481
    42   0.253668  0.313087
    49   0.2 0.2
    52   0.194249  0.145833
    59   0.205252  0.066608 
    67    0.255869 0.0424002
    82    0.343897 0.154636	
    89   0.295481  0.295481
    91   0.253668  0.313087
    98   0.2 0.2
    101   0.194249  0.145833
};
\addlegendentry{d}
\end{axis}
\end{tikzpicture}
\end{subfigure}
\caption{Phase 2}
\label{phase2}
\end{figure}
 The fatigue of distancing impact on contagion because "civic capital", trust in others (section \ref{litreview}) collides with opposite incentives such as social and work pressure generating cycles of huge magnitude for distancing as seen in figure \ref{phase3}\footnote{Case $g=0.04,\mu=0.2,\tau=0.4,\rho=2$ that may pass at 0.25 and 0.5.}.

\begin{figure}[H]
	\centering
	\begin{subfigure}[b]{0.4\textwidth}
\begin{tikzpicture}[scale=0.6]
\begin{axis}[
xlabel=Output,ylabel=Distancing,
title={ Dynamic Paths},
domain=-0:1,
ymax=1,
ymin=0,
xmax=1,
xmin=0,
view={0}{90},
axis background/.style={fill=white},
]
\addplot3 [red, mark=none, thick=3, smooth] table[z=z,x=In,y=dis] {
	
	z	In	dis
	0	0.25	0.5
	0   0.219262  0.593427
	0	0.203324	0.609734
	0   0.178275    0.589615
	0   0.15551     0.50041
	0  0.132741     0.252344
	0  0.141849     0.07770394
	0  0.160064     0.0408351
	0  0.2033324    0.03130077
	0  0.248862     0.0427406
	0  0.276184     0.0884723
	0  0.283015     0.25425
	0  0.25         0.5

};
\addlegendentry{T}

\addplot3[->,red, thick, smooth] 
table[z=z,x=I,y=d] {
	
	z	I	d
	0  0.283015     0.25425	
	0	0.25	0.5
	
};

\addplot3 [
blue,-stealth,samples=15,
quiver={
	u={(0.04*x-0.2*x*y)/(sqrt((0.04*x-0.2*x*y)^2+(-0.4*y+2*x*y)^2))},
	v={(-0.4*y+2*x*y)/(sqrt((0.04*x-0.2*x*y)^2+(-0.4*y+2*x*y)^2))},
	scale arrows=0.05,
},
] {0};
\end{axis}
\end{tikzpicture}

\end{subfigure}
~
\begin{subfigure}[b]{0.4\textwidth}
\begin{tikzpicture}[scale=0.6]
\begin{axis}
[
xlabel=Time,ylabel=Distancing and Output,
title={ Dynamic Paths},
domain=-0:1,
ymax=1,
ymin=0,
xmax=100,
xmin=0,
axis background/.style={fill=white},
]

\addplot [green, mark=none, thick=3, smooth] table[x=z,y=In] {
	
	z	In	dis
	0	0.25	0.5
	8  0.203324	0.609734
	18 0.130464 0.258817
	30 0.15551 0.048801	
	40 0.203324 0.0306266
	50 0.255692 0.048801
	58 0.287568 0.141693
	70 0.203324	0.609734
	78 0.130464 0.258817
	90 0.15551 0.048801	
	100 0.203324 0.0306266

};
\addlegendentry{Y}
\addplot [red, mark=none, thick=3, smooth] table[x=z,y=dis] {
	
	z	In	dis
	0	0.25	0.5
	8 0.203324	0.609734
	18 0.130464 0.258817
	30 0.15551 0.048801
	40 0.203324 0.0306266
	50 0.255692 0.048801
	58 0.287568 0.141693
	70 0.203324	0.609734
	78 0.130464 0.258817
	90 0.15551 0.048801
	100 0.203324 0.0306266		
};
\addlegendentry{d}
\end{axis}
\end{tikzpicture}
\end{subfigure}
\caption{Phase 3}
\label{phase3}
\end{figure}

\begin{figure}[H]  
	\centering
\begin{tikzpicture}
\begin{axis}
[
xlabel=Time,ylabel=Distancing and Output,
title={ Dynamic Paths with phase switches},
domain=-0:1,
ymax=1,
ymin=0,
xmax=305,
xmin=0,
axis background/.style={fill=white},
]
\node[draw] at (50,75) {phase 1};
\node[draw] at (140,50) {phase 2};
\node[draw] at (260,70) {phase 3};
\addplot [green, mark=none, thick=3, smooth] table[x=z,y=In] {
	
	z	In	dis
	0	0.3	0.25
	4	0.253668	0.110622
	7	0.264671	0.0446
	10	0.284478	0.0225939
	20	0.405516	0.00718897
	30	0.557355	0.0424002
	32	0.574971	0.09744178
	35	0.528756	0.214055
	38	0.390112	0.330692
	41	0.3	0.25
	45	0.253668	0.110622
	48	0.264671	0.0446
	51	0.284478	0.0225939
	61	0.405516	0.00718897
	71	0.557355	0.0424002
	73	0.574971	0.09744178
	76	0.528756	0.214055
	79	0.390112	0.330692
	82	0.3	0.25
	86	0.253668	0.110622
	89	0.264671	0.0446
	92	0.284478	0.0225939
	102	0.405516	0.00718897
	103	0.2	0.22
	106	0.194249	0.145833
	113	0.205252	0.066608
	121	0.255869	0.0424002
	136	0.343897	0.154636
	143	0.295481	0.295481
	145	0.253668	0.313087
	152	0.2	0.2
	155	0.194249	0.145833
	162	0.205252	0.066608
	170	0.255869	0.0424002
	185	0.343897	0.154636
	192	0.295481	0.295481
	194	0.253668	0.313087
	201	0.2	0.2
	204	0.194249	0.145833
	205	0.25	0.5
	213	0.203324	0.609734
	223	0.130464	0.258817
	235	0.15551	0.048801
	245	0.203324	0.0306266
	255	0.255692	0.048801
	263	0.287568	0.141693
	275	0.203324	0.609734
	283	0.130464	0.258817
	295	0.15551	0.048801
	305	0.203324	0.0306266

};
\addlegendentry{Y}
\addplot [red, mark=none, thick=3, smooth] table[x=z,y=dis] {
	
	z	In	dis
	0	0.3	0.25
	4	0.253668	0.110622
	7	0.264671	0.0446
	10	0.284478	0.0225939
	20	0.405516	0.00718897
	30	0.557355	0.0424002
	32	0.574971	0.09744178
	35	0.528756	0.214055
	38	0.390112	0.330692
	41	0.3	0.25
	45	0.253668	0.110622
	48	0.264671	0.0446
	51	0.284478	0.0225939
	61	0.405516	0.00718897
	71	0.557355	0.0424002
	73	0.574971	0.09744178
	76	0.528756	0.214055
	79	0.390112	0.330692
	82	0.3	0.25
	86	0.253668	0.110622
	89	0.264671	0.0446
	92	0.284478	0.0225939
	102	0.405516	0.00718897
	103	0.2	0.22
	106	0.194249	0.145833
	113	0.205252	0.066608
	121	0.255869	0.0424002
	136	0.343897	0.154636
	143	0.295481	0.295481
	145	0.253668	0.313087
	152	0.2	0.2
	155	0.194249	0.145833
	162	0.205252	0.066608
	170	0.255869	0.0424002
	185	0.343897	0.154636
	192	0.295481	0.295481
	194	0.253668	0.313087
	201	0.2	0.2
	204	0.194249	0.145833
	205	0.25	0.5
	213	0.203324	0.609734
	223	0.130464	0.258817
	235	0.15551	0.048801
	245	0.203324	0.0306266
	255	0.255692	0.048801
	263	0.287568	0.141693
	275	0.203324	0.609734
	283	0.130464	0.258817
	295	0.15551	0.048801
	305	0.203324	0.0306266

};
\addlegendentry{d}
\end{axis}

\end{tikzpicture}
\caption{Trajectories of $Y$ and $d$}
\label{Phases123}

\end{figure}
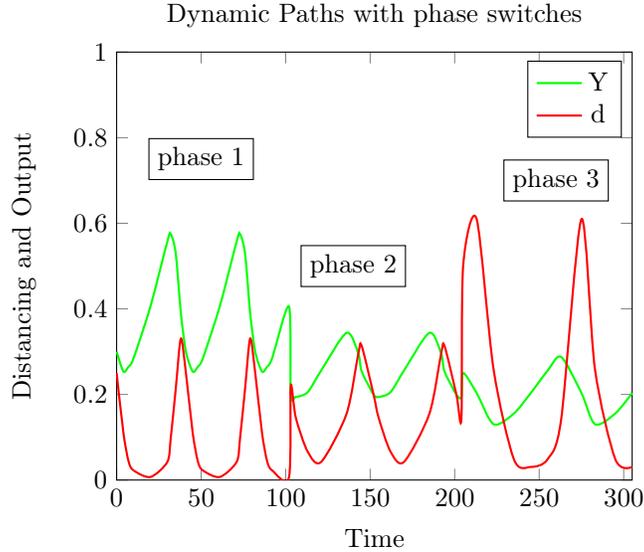

Figure \ref{Phases123}: summarizes the paths of figures \ref{phase1},\ref{phase2},\ref{phase3};

Following the sequence of the figures (\ref{phase1}; \ref{phase2}; \ref{phase3}) as phases od the dynamic process we can see that initially the impact of the contagion finds a first phase where the level of activity continues to exceed the distancing (phase 1). However, the increase in infections forces the system to enter a new phase (phase 2), where the level of activity begins to fluctuate in a lower range. As restrictions are imposed (phase 3), activity continues at low levels, while greater distancing restrictions are imposed. 

The example of figure \ref{Phases123} shows the dependency on the susceptibility parameter $\rho$.  Even when we see $\mu$ falling (say, due to that the economy is adapting to a mode of production with remote working) and $\tau$ is also decreasing, given the adaptation of factor $N$; the cyclical behavior of $\rho$ force the system to a cycle with lower output $Y$ with high fluctuations in social distancing $d$.

The dynamics presented in figure \ref{Phases123} is similar to the path  presented by countries like Germany with a first peak of a little more than 5000 in March-April (2020) following a long valley (cases in the hundreds) reaching more than 25000 in November 2020 with a valley of near 7000 cases in February 2021 and a new high of more than 20000 in April 2021. For Argentina the first peak was near 15000 cases in October 2020 followed by a valley with values as low as 5000 and intermediate peaks of near 12000 reaching near 25000 in April 2021 for contagion for the 7 days average \cite{hopkins2021}. Following waves of contagion, social distancing is widely implemented.

Similar paths (tendencies) for GDP to the one presented in figure \ref{Phases123} are reported in empirical evidence. Worldwide, GDP presents a huge fall in the first halve of 2020, while the recovery that followed was not enough to reestablish previous levels. In the G20 area GDP rebounded 8.1\% in the third quarter of 2020, while remaining 2.4\% lower than pre-crisis levels of the last quarter of 2019 \cite{2020PBI}.

\section{Conclusions}

In the model of gravitation created by Isaac Newton, the force of gravity requires a significant counterforce in order to escape its bounds, as, for example, to launch a rocket from Earth to outer space. If that counterforce were reduced beyond the threshold needed for escape velocity, the rocket would once again come under the influence of gravity and fall back to Earth. If the counterforce is somehow recovered, the rocket would raise again, generating trajectories with cyclical patterns. 

The COVID-19 pandemic and the manner in which it is spread has presented economic agents with extremely unpleasant choices regarding their behavior – in many cases, pushed along by government mandates - leading to direct effects in the real economy. In particular, the prophylactic of social distancing may have been a useful mechanism by which to slow the spread of the disease, and, if it actually is effective in limiting contagion, would yield long-term benefits in preserving the work force and hastening an end to the pandemic. On the other hand, the restrictions of social distancing had more immediate and likely medium-term effects, namely the opportunity cost of lower economic activity and output and possible follow-on effects in lower investment for the future. To come back to Newton and our space-bound rocket, the pandemic has shown us that social distancing forces, used to reduce contagion, were counterbalanced by the resistance imposed by the organization of production and the institutions created and sustained by human interaction. It is a “social gravitational force” which resists distancing and is necessary for production, and the force/counterforce interplay creates different trajectories depending upon which force is favored.

This paper models these trade-offs in the context of an economy, integrating various parameter configurations into a succession of phases associated with social distancing. Our key result is that the trade-off of lower output in the long run for defeating the disease via social distancing cannot be avoided: lower contagion means lower output ceteris paribus.  One may defy gravity for a short time, but there are consequences. Indeed, this economic reality, that various courses of actions are mutually exclusive trade-offs, is the motivating factor behind the key point of our model, namely that it does not have an explicit singular solution. Instead, the model is offered as an attempt to illuminate some of the variables and parameters which may be significant in the context of the optimization problem. However, this also highlights one of the limitations of the model, as well as of public policy in response to a pandemic in general. As the model depends on a specific set of parameters, $g$ Gdp growth, $\mu$ economic structure and policies, $\tau$ response of the labor force to distancing and $\rho$ the susceptibility of the system to the need of more social distancing $d$, our simulations show the complex nature of the casual relations between the variables. Even in a sterile, disease-free, simulation setting, we may not be able to provide the exact relations and responses, as, they are highly path dependent: much like the meltwater under a glacier in Greenland \cite{poinar2015limits}, which may deposit quietly in a lake or gain force, melt more ice, and form a river to the ocean, we do not know which force will predominate and determine the reaction of the system. 

Despite this limitation, we hope to have provided a simulation tool as a first attempt to model these relationships explicitly and understand the extent of the trade-offs in setting policy. Indeed, the limitations of the model also provide a warning for policymakers, as the complexity of pandemic response means that there are no simple solutions, but rather, difficult trade-offs.  Moreover, our findings suggest that the interaction of forces at play during the pandemic were non-linear, introducing additional complexity. As a policymaker, veering from policy to policy and influenced by the ecosystem of institutions surrounding policymaking, trade-offs and non-linearities may not be readily apparent but must be considered and, importantly, communicated. Pandemic response is not a binary “health or the economy” choice, but instead a continuum where there are opportunity costs in the short-, medium-, and long-term. To pretend otherwise is folly and, most likely, counterproductive for building the trust necessary to overcome such an exogenous shock.

\appendix
\section{Appendix: A non-consistent case}\label{saddle}

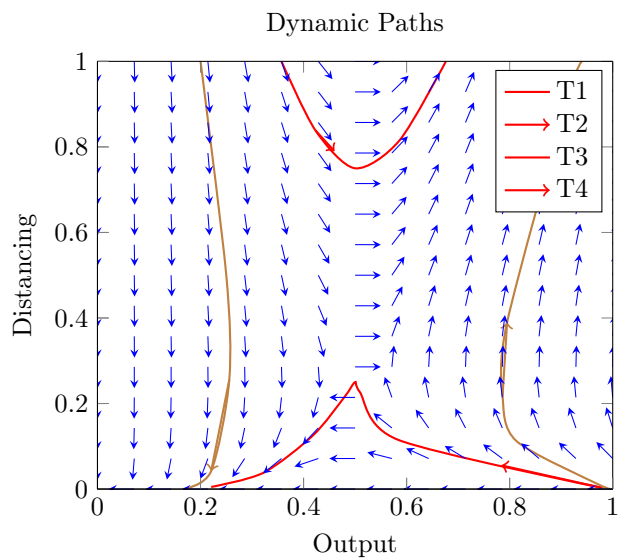
\begin{figure}[H]
	\centering
	
	\begin{tikzpicture}
	\begin{axis}[
	xlabel=Output,ylabel=Distancing,
	title={ Dynamic Paths},
	domain=-0:1,
	ymax=1,
	ymin=0,
	xmax=1,
	xmin=0,
	view={0}{90},
	axis background/.style={fill=white},
	]
	\addplot3 [red, mark=none, thick=3, smooth] table[z=z,x=In,y=dis] {
		
		z	In	dis
		0 0.98 0.00278756
		15 0.583773 0.115023
		30 0.50895  0.227259
		50 0.5  0.25
		59 0.48 0.216256
		70 0.418721 0.126027
		80 0.32189 0.0379988
		90 0.220657 0.00498826

	};
	\addlegendentry{T1}

	\addplot3[->,red, thick, smooth] 
	table[z=z,x=I,y=d] {
		
		z	I	d
		0 0.98 0.00278756
		15 0.783773 0.0555023
		
	};
	
	\addplot3 [red, mark=none, thick=3, smooth] table[z=z,x=In,y=dis] {
		
		z	In	dis
		0  0.357101 1
		2  0.423122 0.83844
		4  0.5 0.75
		6 0.585974  0.814847
		7.5 0.676202 1
		
	};
	\addlegendentry{T2}
	
	\addplot3[->,red, thick, smooth] 
	table[z=z,x=I,y=d] {
		
		z	I	d
		2  0.423122 0.83844
		4  0.46 0.79

	};
	
	\addplot3 [brown, mark=none, thick=3, smooth] table[z=z,x=In,y=dis] {
		
		z	In	dis
		0 0.2 1
		3 0.25 0.5
		6 0.255869 0.255869
		16 0.220657 0.042002
		24 0.16564 0
		
	};
	\addlegendentry{T3}
	
	\addplot3[->,brown, thick, smooth] 
	table[z=z,x=I,y=d] {
		
		z	I	d
		6 0.255869 0.255869
		16 0.220657 0.042002

	};
	
	\addplot3 [brown, mark=none, thick=3, smooth] table[z=z,x=In,y=dis] {
		
		z	In	dis
		0 1 0
		2 0.971097 0.0115904
		8 0.812646 0.112823
		10 0.786238 0.218457
		12 0.795041 0.390112
		14.61 0.911678 0.911678
		14.9 0.940288 1

	};
	\addlegendentry{T4}
	
	\addplot3[->,brown, thick, smooth] 
	table[z=z,x=I,y=d] {
		
		z	I	d
		10 0.786238 0.218457
		12 0.795041 0.390112

	};
	
	\addplot3 [
	blue,-stealth,samples=15,
	quiver={
		u={(-0.04*x-(-0.15)*x*y)/(sqrt((-0.04*x-(-0.15)*x*y)^2+(-0.5*y+1*x*y)^2))},
		v={(-0.5*y+1*x*y)/(sqrt((-0.04*x-(-0.15)*x*y)^2+(-0.5*y+1*x*y)^2))},
		scale arrows=0.05,
	},
	] {0};
	\end{axis}
	\end{tikzpicture}
	
	\caption{Non consistent cases}
	\label{nonconsitent}
\end{figure}

Economic consistency may help us to discard some cases. We may dismiss the case with negative income grow and a positive influence of distancing over the growth rate (think of an economy based on logistic distribution or strong software capabilities)\footnote{For example, case saddle at Y=0.5 and d=0.25, with $g=-0.04,\mu=-0.15,\tau=0.5,\rho=1$.}. We present several theoretical possible trajectories in figure \ref{nonconsitent}.

\bibliographystyle{apacite}
\bibliography{rnb} 
\end{document}